# (R)evolution of Programming: Vibe Coding as a Post-Coding Paradigm

Kevin Krings*

University of Hagen, EduTech for Digital Transformation, 58097 Hagen, Germany, kevin.krings@fernuni-hagen.de

Nino S. Bohn

University of Hagen, EduTech for Digital Transformation, 58097 Hagen, Germany, nino.bohn@fernuni-hagen.de

Thomas Ludwig

University of Hagen, EduTech for Digital Transformation, 58097 Hagen, Germany, thomas.ludwig@fernuni-hagen.de

Recent advancements in generative artificial intelligence (GenAI), particularly large language models, have introduced new possibilities for software development practices. In our paper we investigate the emerging Vibe Coding (VC) paradigm that emphasizes intuitive, affect-driven, and improvisational interactions between developers and AI systems. Building upon the discourse of End-User Development (EUD), we explore how VC diverges from conventional programming approaches such as those supported by tools like GitHub Copilot. Through five semi-structured interview sessions with ten experienced software practitioners, we identify five thematic dimensions: creativity, sustainability, the future of programming, collaboration, and criticism. Our analysis conceptualizes VC within the metaphor of co-drifting, contrasting it with the prevalent co-piloting perspective of AI-assisted development. We argue that VC reconfigures the developer's role, blurring boundaries between professional and non-developers. While VC enables novel forms of expression and rapid prototyping, it also introduces challenges regarding reproducibility, scalability, and inclusivity. We propose that VC represents a meaningful shift in programming culture, warranting further investigation within human-computer interaction (HCI) and software engineering research.

CCS CONCEPTS • **Software and its engineering ~ Software creation and management ~ Software development techniques ~ Software prototyping** • Software and its engineering ~ Software creation and management ~ Software development process management ~ Software development methods ~ Rapid application development • Software and its engineering ~ Software creation and management ~ Designing software ~ Software implementation planning ~ Software design techniques • **Human-centered computing ~ Human computer interaction (HCI) ~ Empirical studies in HCI** • **Human-centered computing ~ Human computer interaction (HCI) ~ Interaction paradigms ~ Natural language interfaces**

---

\* Corresponding Author.

# 1 INTRODUCTION AND BACKGROUND

In the past, software engineering has demanded specialized (technical) expertise [5], making it unapproachable to those without programming skills. While the shift from command-line interfaces to graphical integrated development environments (IDEs) and the shift from low-level to high-level programming languages marked early steps toward improving usability [5, 22, 25], coding skills remained mandatory. This challenge is focused by the established research discourse on End-user development (EUD) [1, 15, 16, 19, 20] and by the empowerment of non-programmers to adapt technologies to their own practices and vice versa [4]. Specifically, EUD describes the sub-process of adapting new technologies to users' practices and the emergence of new practices in situ [4, 17, 24]. Lieberman et al. [15] define EUD as a collection of methods, techniques, and tools that enable non-developers to create, adapt, and extend software, while approaches that focus more on user practices [16] refer to EUD when non-developers need to transition to a more abstract level to perform their specific tasks. Conventional applications often present steep learning curves for end-users, requiring a high level of knowledge to customize the program to their liking [16]. Paternò [19] characterizes these steep learning curves of conventional applications as walls that users must stop at and learn many new concepts and techniques before they are able to step forward. Therefore, EUD aims to provide applications that offer a gradual increase in complexity, resulting in a shallower learning curve for end-users [16]. Approaches aim to reduce complexity through model-based development [15], assistant-based approaches [21], or simplifying abstraction of various functionalities [18] to support non-developers. This includes creating flatter hierarchies, limited functionalities, avoiding multiple ways to achieve the same function, using sliders, drag-and-drop approaches, employing Natural User Interfaces (NUI), or incorporate generative artificial intelligence (GenAI) [6, 11, 12, 18, 19].

Recent breakthroughs in GenAI, especially in the area of large language models (LLMs) are beginning to reshape this landscape by allowing software to be created through more intuitive, natural language interactions. The ongoing shifts in programming practices, like co-coding as highlighted by Welsch [26] and Jonsson & Tholander [9], are closely tied to the rise of new paradigms such as Vibe Coding (VC). This approach reflects a broader trend where programmers are increasingly required to transfer some control over the coding and creative processes. Andrej Karpathy describes it as „*[...] a new kind of coding [...], where you fully give in to the vibes, embrace exponentials, and forget that the code even exists. [...] I barely even touch the keyboard. I ask for the dumbest things [...], because I'm too lazy to find it. I ‚Accept All' always, I don't read the diffs anymore. When I get error messages I just copy paste them in with no comment, usually that fixes it. The code grows beyond my usual comprehension. I'd have to really read through it for a while. Sometimes the LLMs can't fix a bug so I just work around it or ask for random changes until it goes away. It's not too bad for throwaway weekend projects, [...] but it's not really coding [...].* [10]" VC challenges traditional programming by prioritizing fluid, intuitive, and spontaneous code generation and leads to rapid prototyping and experimental outputs. Unlike tools such as GitHub Copilot, which assist developers by generating code snippets based on textual prompts, VC embraces improvisation and creative expression. This shift from structured to emergent coding practices raises questions about sustainability and practicality. While Copilot primarily augments existing workflows by enhancing efficiency, VC seeks to break free from conventional code logic, aiming for a more human-centered coding experience. Due to its novelty, VC is rarely discussed in academia till now. Chow and Ng [3] for example applied VC to develop two medical applications, demonstrating improvements in accessibility, cost-effectiveness, and scalability. Sapkota et al. [23] focus on the comparison of VC and agentic coding to provide a taxonomy covering theoretical and conceptual foundations, execution paradigms, feedback loops, safety mechanisms, debugging methodologies, and practical toolchains.

This paper aims to examine VC as a potential alternative to traditional programming paradigms, considering the efforts within the EUD research discourse, with a particular focus on its differences from AI-supported coding tools like GitHub Copilot. By drawing insights from five semi-structured interviews and analyzing (critical) perspectives, we aim to answer *how VC differs from traditional AI-assisted programming approaches* and situate the potential of VC, as a post-coding paradigm, in reshaping programming practices.

While EUD traditionally focuses on non-programmers, we engaged professional developers in this study to understand how emerging practices, such as VC, may shift foundational boundaries of programming expertise itself. Rather than viewing developers as a distinct group apart from end users, we approach them as early adopters and boundary actors exploring hybrid roles. In these roles, creative expression, improvisation, and interaction with AI tools begin to blur the lines between professional programming and end-user activity, reflecting a shift already observed in a previous study on novel EUD paradigms [11]. By examining how experienced coders themselves grapple with loss of control, improvisation, and non-traditional workflows, we gain insights into how programming might become more accessible, expressive, and resonant for a broader range of users in the future.

## 2 INTERVIEW STUDY

To gain insights into the use and the potential of VC, we conducted five semi-structured interview sessions with ten experienced practitioners and software developers in total. The interviews focused on personal experiences, practices, and reflections related to AI-assisted coding, without introducing predefined categories. Participants were encouraged to describe workflows, tools, moments of uncertainty, and their own interpretations of what it means to program using this approach. The interview sessions were conducted in Q2 2025 and had an average duration of 43 minutes. They were conducted with one to four people at the same time. The participants were between 29 to 56 years old, while there were nine male and one female. Through purposive sampling, we recruited software developers from various industries, all of whom had experience integrating AI tools into their development processes [7]. We conducted a thematic analysis following Braun and Clarke [2] to investigate the research question. Thematic coding revealed five major themes.

### 2.1 Results

*2.1.1 Theme 1: Creativity*

VC is described as an intuitive, vibe-driven practice where *"code emerges from mood"* and *"you could just try out random stuff until something seems to work the way you want it to work."* Unlike tools like Copilot, which support functional efficiency, VC prioritizes affective expression. It fosters creativity and improvisation by allowing spontaneous code generation without the burden of syntactical rigor. One participant explained, *"Vibe Coding lets you express ideas without worrying about the technicalities"* as you can *"see stuff, say stuff, run stuff, and copy and paste stuff and it mostly works."* In contrast, assistive tools like Copilot focus on code completion rather than spontaneous ideation, limiting creative coding flow.

*2.1.2 Theme 2: Sustainability*

A significant critique of VC is its tendency to produce short-lived, disposable software. One interviewee noted, *"You create something that works in the moment, but it's rarely usable beyond that."* While tools like Copilot are integrated into IDEs, VC's experimental nature currently leads to issues of scalability and long-term usability.

*2.1.3 Theme 3: Future of Programming*

Some participants see VC as a paradigm shift, while others question its long-term viability. A developer noted, *"it's exciting as a concept, but in practice, it often lacks robustness."* Unlike integrated co-coding approaches, which fits seamlessly into conventional software development pipelines, VC challenges the notion of coding as a structured and predictable task [26]. Furthermore, a participant remarked that *"Copilot finishes my thoughts, Vibe Coding changes them."* This destabilization raises critical questions about agency, authorship, and unpredictability. It is predicted that *"the role of software engineer will transition to product engineer"* because *"human taste is now more important than ever as VC tools make everyone a 10x engineer."*

*2.1.4 Theme 4: Collaboration*

VC shows potential for creative teamwork, as it encourages collaborative ideation and rapid prototyping. However, reproducibility and consistency are significant challenges. One participant noted, *"It's hard to build on someone else's vibes. It feels too personal and context-specific."* In contrast, tools such as Copilot support collaboration through shared codebases, while VC remains inherently individualistic.

*2.1.5 Theme 5: Criticism*

Practical concerns dominate the discourse around VC, including the difficulty of adapting vibe-coded projects for other use cases. A participant remarked, *"Vibe Coding is great for getting ideas out fast, but you often end up with code that's a dead end."* Furthermore, there are insufficient code validation opportunities and overreliance on AI-generated outputs as one *"can accidentally ship bugs that are so complex... a security expert and two engineers didn't see the problem. It was found by accident."* Additionally, the financial burden of using AI tools for spontaneous coding is a recurring issue. Our participants highlighted its reliance on API credits and powerful machines, raising concerns of exclusion and tech elitism. Integrated co-coding such as Copilot reduce such costs and aligns more with production needs.

## 3 DISCUSSION

In current narratives around AI-assisted programming, the *co-piloting* metaphor has become dominant. This frames the AI as a helpful assistant in a clearly goal-directed task. The relationship is centered around efficiency, predictability, and control. The collaboration is productive and bounded, aiming at better code, faster delivery, and fewer errors. Our study around VC points to an alternative metaphor which we call *co-drifting* (see Table 1). It describes a vibe-driven practice where *"code emerges from mood."* This resonates with science fiction depictions of programming through body movement or ritualistic interaction [11]. At the same time it *"lets you express ideas without worrying about the technicalities"*, what correspond with Lewis [14] suggesting that GenAI can facilitate creative thinking by reducing cognitive barriers.

Where Copilot aligns with industrial workflows, VC disrupts them. In contrast to the structured guidance of a *co-piloting* AI, *co-drifting* suggests a more open-ended, improvisational dynamic between human and AI. Here, neither party leads, instead, both "drift" together through the creative process. The emphasis shifts from control to resonance, from task completion to emergent exploration.

Table 1: Comparison between co-piloting and co-drifting metaphors

| Aspect | Co-Piloting | Co-Drifting |
| --- | --- | --- |
| Control | Human leads, AI assists | Human and AI drift together |
| Objective | Efficiency, accuracy, productivity | Exploration, emergence, surprise |
| Orientation | Task- and goal-oriented | Process- and experience-oriented |
| Output | Structured, reusable code | Ephemeral, expressive traces |
| Relation to AI | Guided collaboration | Open-ended resonance |

| Aspect | Co-Piloting | Co-Drifting |
|---|---|---|
| Role of User | Strategic planner | Creative co-creator |
| Tool examples | GitHub Copilot, TabNine | Cursor, v0 (Vercel), bolt |

Our participants underscore the transient nature of VC as *"you don't save the script, you save the flow"* and that one *"don't write code much. I just think and review."* This position code less as durable infrastructure and more as creative residue, opens questions about value, waste, and memory. While some view this as a breakthrough in co-creative programming, comparable to already identified AI-enhanced programming paradigms such as *programming by argumentation* [11], others argue that it may be unsustainable due to its high costs and the creation of transient, disposable software, suggesting conceiving VC as a post-coding paradigm within the EUD research discourse. This reframing pushes EUD into a new territory. Till now, EUD approaches focuses on empowering non-programmers by flattening learning curves, simplifying abstractions, or hiding complexity [6, 11, 12, 18, 19]. *Co-Drifting*, in contrast, does not simplify the system, it changes the relationship to it. Professional developers adopt the posture of an end-user: exploratory, intuitive, occasionally passive. From this perspective, VC is a form of expressive EUD for experts and non-developers, a way to reinhabit their tools through improvisation, surprise, and even aesthetic pleasure. The practice dissolves the clear dichotomy between developer and user.

The reframing from *co-piloting* to *co-drifting* opens up space for speculative programming imaginaries that value process over product, ambiguity over precision, and mutual transformation over optimization. In this light, we argue that VC and *co-drifting* should be taken seriously, not as a fringe phenomenon or hobbyist indulgence, but as a critical lens for rethinking programming itself in the age of GenAI.

## 4 CONCLUSION

VC challenges established programming paradigms by blending creative spontaneity with AI-driven code generation. While it holds promise for innovative co-creation, its ephemeral nature and cost-intensive setup raise critical questions about sustainability, scalability, and practical utility. By comparing VC with structured AI-assisted coding tools like Copilot, we aim to illuminate new research directions in HCI.VC invites us to rethink programming as an expressive and speculative practice. When juxtaposed with science fiction imaginaries [11], it reveals a rich design spaces beyond productivity and precision focusing on cultural critique, experimentation, and alternative futures. In this context, VC aligns with the concept of otherware [8, 13], that implies a fundamentally different relationship between human and technology. Whether by design or coincidence, otherware manifests as a computational other, rather than as an embodied projection of the self [13]. We argue that VC is not just a quirky trend, but science fiction in action, one that demands serious engagement from the HCI community.

**ACKNOWLEDGMENTS**

We would like to thank all participants who took part in our study. Additionally, we thank our collective mood and the vibes within our research group, which oscillated between existential dread and euphoric insight.